\documentclass[aps,twocolumn,superscriptaddress,showpacs,amsmath,amssymb]{revtex4}

\usepackage{mathrsfs}
\usepackage{txfonts}

\usepackage{graphicx}
\usepackage{dcolumn}% Align table columns on decimal point
\usepackage{bm}% bold math

\begin{document}
\title{Phenomenology analysis of duration inflation for Tachyon field in loop quantum cosmology}
\author{Kui Xiao}
 \email{87xiaokui@mail.bnu.edu.cn}
  \affiliation{Department of Mathematical and Physical Teaching, Hunan Institute of Technology, Hengyang 421002, China}
\author{Xiao-Kai He}
 \email{hexiaokai77@163.com}
  \affiliation{{Department of Educational Science}, Hunan First Normal University, Changsha 410205, China}
\author{Fei Huang}
 \email{huangfei@email.arizona.edu}
 \affiliation{Department of Physics, University of Arizona,1118 E. 4th St., Tucson, AZ 85721, USA}
\author{Jian-Yang Zhu}
\thanks{Corresponding author}
 \email{zhujy@bnu.edu.cn}
  \affiliation{Department of Physics, Beijing Normal University, Beijing 100875, China}
\date{\today}
\begin{abstract}
Assuming that the e-folding number is just determined by the change of the scale factor, the tachyonic inflation theory in LQC has been discussed. Considering the tachyon field with exponential potential and inverse quadratic potential, we find that the evolutionary pictures of super inflation are affected by the potentials and the initial conditions. However it cannot provide enough e-folding number, no matter which condition is chosen. Therefore a slow-rolling inflation is necessary. The e-folding number for slow-rolling inflation depends on the values of the parameter $\alpha$ of the exponential potential and the initial conditions. To get enough e-folding number, $\alpha$ should be small. Based on the slow-rolling inflation happens immediately when the super inflation ends, and the scale factor is continuously growing during the whole inflation stage, we consider an e-folding number provided by the whole inflationary stage, and we find that it is easier to get enough e-folding number when the scale factor increases during all the inflation phase.
\end{abstract}

\pacs{98.80.Cq}
\maketitle

\section{Introduction}

Recently, researchers from the BICEP2 collaboration announced the first direct evidence for the cosmic inflation \cite{BICEP2}. The inflation theory is introduced to solve lots of cosmological conundrums (the monopole, horizon, flatness, and entropy problems) in the standard cosmological model \cite{Lindle}. Inflationary models have has made huge success. In particular, it generates fluctuations that became the seeds for the growth of structures \cite{Weinberg}.  Inflation happens while the Hubble parameter $H$ is  approximately a constant, and $\ddot{a}>0$. In this stage, the equation of state (EoS) parameter of inflationary field is $\omega=-1$ which means the potential energy of inflationary field is dominant. This domination continues until $\omega=-1/3$, which indicates the kinetic energy can no longer be ignored. Lots of evidences show that inflation is a brilliant candidate theory to explain the very early universe (i.e., see \cite{inflation-planck,Lisa-warm}).  But there are still some problems need to be solved, i.e., the generating initial conditions for inflation, the singularity of the universe, and so on. As we know, if the universe is filled  with  radiation and matter, according to the general relativity theory, coming back in time, one concludes  that there exists a primeval singularity, i.e., big bang singularity. The big bang singularity problem could be viewed as a defect of Einstein cosmology at high energies scale.

To solve the big bang singularity problem, one possible solution is modifying the theory of general relativity at high energy scale. There are lots of candidates, one of which is LQC (more recently review, please see \cite{Ivan-LQC,Ashtekar-re,Bojowald-book}). LQC  is a canonical quantization of homogeneous spacetimes based upon the techniques used in loop quantum gravity (LQG)\cite{Rovelli-Book,Thiemann-Book}. Due to the homogeneity and isotropy of spacetime, the phase space of LQC is simpler than LQG, e.g., the connection and triad are described by just two quantities $c$ and $p$, respectively. The dynamics of LQC can be studied effectively by introducing quantum gravity corrections to the gravity and matter Hamiltonian \cite{Ivan-LQC,Bojowald-book}, and the numerical evidence shows that the effective equations provide an excellent approximation to the full dynamics of sharply peaked states \cite{Ashtekar-74,Corichi,Diener-1,Rovelli,Diener}. In general,  two types of corrections are considered:  inverse volume correction and holonomy correction. Considering these modification, one can obtain many interesting results, e.g., the replacement of big bang by big bounce \cite{Ashtekar-74,Ashtekar}, the avoidance of most singularities \cite{Singh}, the more likely occurrence of inflation\cite{Singh-in,Ashtekar-probability-1,Ashtekar-probability,Corichi_measure,Linsefors,Singh-in-2}, and so on. But the first modification suffers from gauge dependence which cannot be cured and thus yields unphysical effects. Therefore we will discuss the tachyonic inflation theory in LQC based on the second modification. In this effective LQC, a term of $-\rho^2/\rho_c$ will be added to the right hands of the standard Friedmann equation. Since this correction comes with a negative sign, the Hubble parameter $H$,  then $\dot{a}$ will vanishes when $\rho=\rho_c$, at which the quantum bounce occurs and the universe oscillates forever.

Since the Friedmman equation adds a term of $-\rho^2/\rho_c$ in effective LQC, the universe will enter a super inflation stage as soon as the quantum bounce happens, i.e., $\rho=\rho_c$. The super inflation is totally caused by the quantum geometry effect in LQC \cite{Singh-PRD}, and it has been studied by many papers \cite{Singh-PRD,Copeland-superinflation,Ranken,Ribassin,Amoros-prd,Xiao-Superinflation}. The generating initial conditions for inflation are far from easy in Einstein cosmology, but there are lots of works argued that the phase of inflation is generic for the model with the holonomy corrections in LQC \cite{Singh-in,Ashtekar-probability-1,Ashtekar-probability,Corichi_measure,Linsefors,Singh-in-2}. The inflation theory in LQC has been discussed by many papers and the observational issues of LQC also have been studied (for more recently review, please see \cite{Barrau}).

We will study the tachyonic inflation in this paper. Tachyon field might be responsible for the inflation at the early stage and could contribute to some new form of dark matter at late time \cite{Sen-JHEP}. The behavior of tachyon field in LQC was studied in \cite{Sen-PRD}, in which they considered the inverse volume modification. They found that there exists a super accelerated phase in the semiclassical region. The dynamical behaviors of non-interacting and interacting tachyon field are studied in \cite{Huang,Tavakoli}. It was shown there, the dynamical behavior of tachyon field is different from the one in Einstein cosmology. The inflation theory of tachyon field in LQC also been discussed \cite{Xiong,Xiao,Wu}. The evolution picture and inflation of tachyon field with exponential potential in LQC was studied in \cite{Xiong}. It was shown there, for small value of $\dot{\phi}$ the quantum trajectories approach the classical inflationary attractor. But they just considered the inflation stage, and the super-inflation is ignored.  The purpose of this paper is to discuss how to get enough e-folding number of inflation in LQC, we will discuss the super inflation theory of tachyon field with exponential potential and inverse square potential in LQC, and the slow-roll inflation of tachyon field with exponential potential in LQC will be discussed, and we will show the enough e-folding number will be easier to get when one considers the super inflation and slow-roll inflation as a unitary inflation stage.

This paper is  organized as follows. Firstly, we will introduce some basic equations of LQC and the tachyon field in section \ref{II}, and then, in section \ref{III}, we will show the tachyonic inflation in LQC. Finally, we draw the conclusions in section in the last section. For simplicity, we set $8\pi G=1$.

\section{Framework\label{II}}

We focus on the flat FRW cosmology in this paper. The effective equations in LQC are derived from the LQC hamiltonian constraint operator and  include  the leading order quantum gravity
corrections to the classical Friedmann equation. It turns out that the effective equations provide a surprisingly good approximation to the dynamics of sharply peaked states in LQC at all times, including at the bounce point where quantum gravity effects are strongest \cite{Ashtekar-74,Ashtekar,Diener}. In this paper, we focus on the holonomy correction, then the modified Friedmann equation reads
\begin{eqnarray}
H^2=\frac{1}{3}\rho\left(1-\frac{\rho}{\rho_c}\right), \label{Fri}
\end{eqnarray}
in which $\rho_c$ is the critical energy density. For such model, energy density of matter cannot exceed the critical energy density. We consider the universe is sourced by a tachyon field. The energy density and the pressure of tachyon field are
\begin{eqnarray}
\rho=\frac{V}{\sqrt{1-\dot{\phi}^2}},\qquad p=-V\sqrt{1-\dot{\phi}^2},\label{rhop}
\end{eqnarray}
in which $V$ is the potential of tachyon field.  The exponential potential $V=V_0 e^{-\lambda \phi}$ with constants $V_0,\lambda$, the quadratic potential $V=\frac12\sigma \phi^2$ with constant $\sigma$, and inverse quadratic potential $V=-\beta \phi^{-2}$ with constant $\beta$ are always discussed when the inflation properties of tachyon field are studied (see, \emph{i.e.}, \cite{Nozari}). We will consider the inverse quadratic potential and exponential one in this paper.

Using Eq. (\ref{rhop}), and considering the continuity equation $\dot{\rho}+3H(\rho+p)=0$, one can get the evolution equation of the tachyon field as
\begin{eqnarray}
\ddot{\phi}+(1-\dot{\phi})\left(3H\dot{\phi}+\frac{V'}{V}\right)\label{EoM},
\end{eqnarray}
where prime denotes differentiation w.r.t. $\phi$.

Considering the Eqs. (\ref{Fri}) and (\ref{rhop}) and the continuity equation, one can get
\begin{eqnarray}
\dot{H}=-\frac12(\rho+p)\left(1-\frac{2\rho}{\rho_c}\right).\label{dH}
\end{eqnarray}
According to Eq.(\ref{Fri}), we can find that $H=0$ when $\rho=\rho_c$, which is the quantum bounce point. It is easy to see that $\dot{H}>0$ at the bounce point  and it holds
positive  until $\rho=\frac12\rho_c$. $\dot{H}>0$ means the universe is in a super inflation stage. The Hubble parameter and the scale factor increase during this stage. And since super inflation is completely due to the modification term in the Friedmann equation, it is rooted in the quantum geometry effect.

Using Eqs.(\ref{Fri}) and (\ref{dH}), it is easy to get the modified Raychaudhuri equation
\begin{eqnarray}
\frac{\ddot{a}}{a}=\dot{H}+H^2=\frac{1}{3}\rho\left(1-\frac{\rho}{\rho_c}\right)-\frac12(\rho+p)\left(1-\frac{2\rho}{\rho_c}\right).\label{dda}
\end{eqnarray}

\begin{figure}[!ht]
\includegraphics[width=0.48\textwidth]{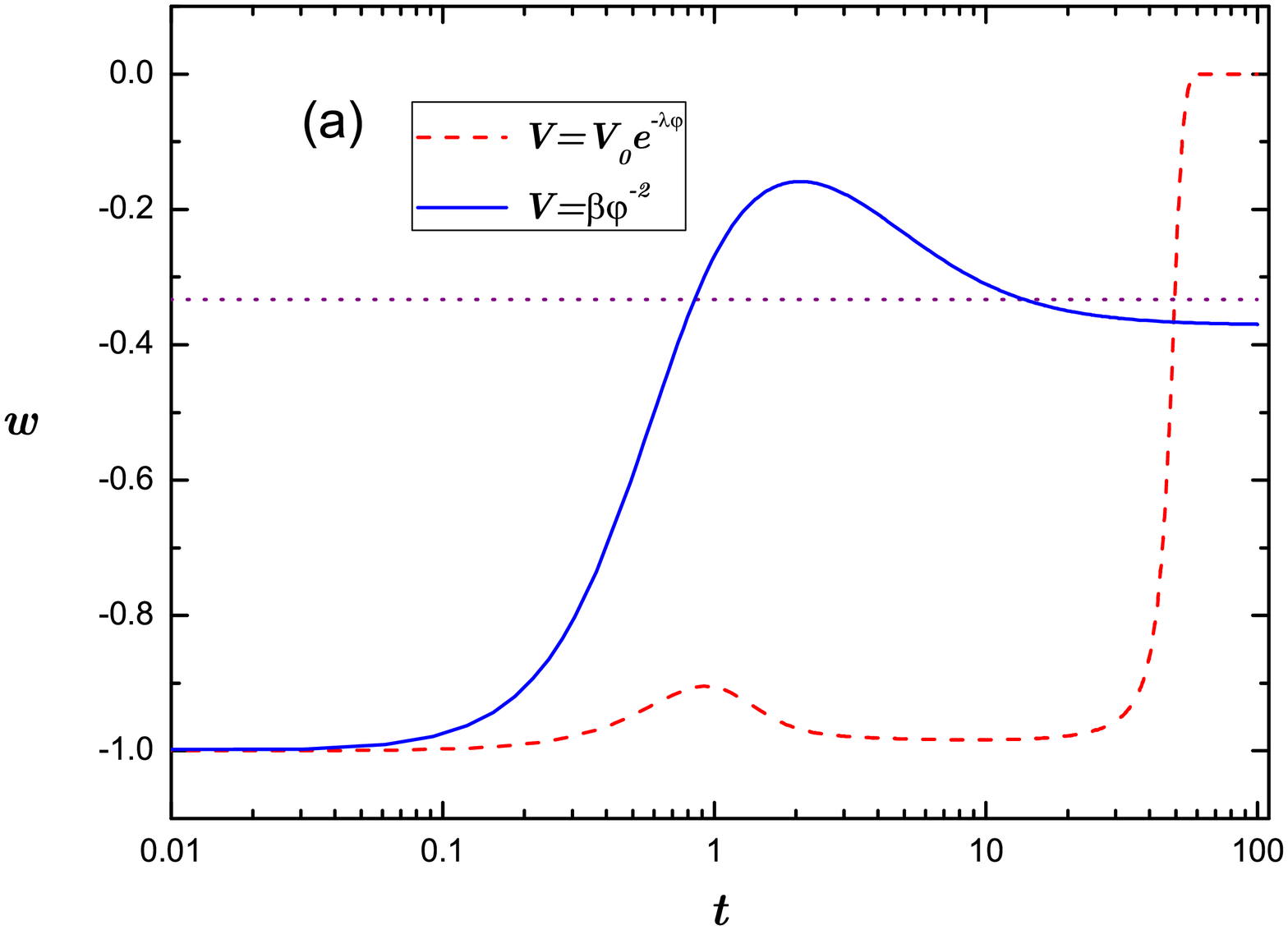}
\includegraphics[width=0.48\textwidth]{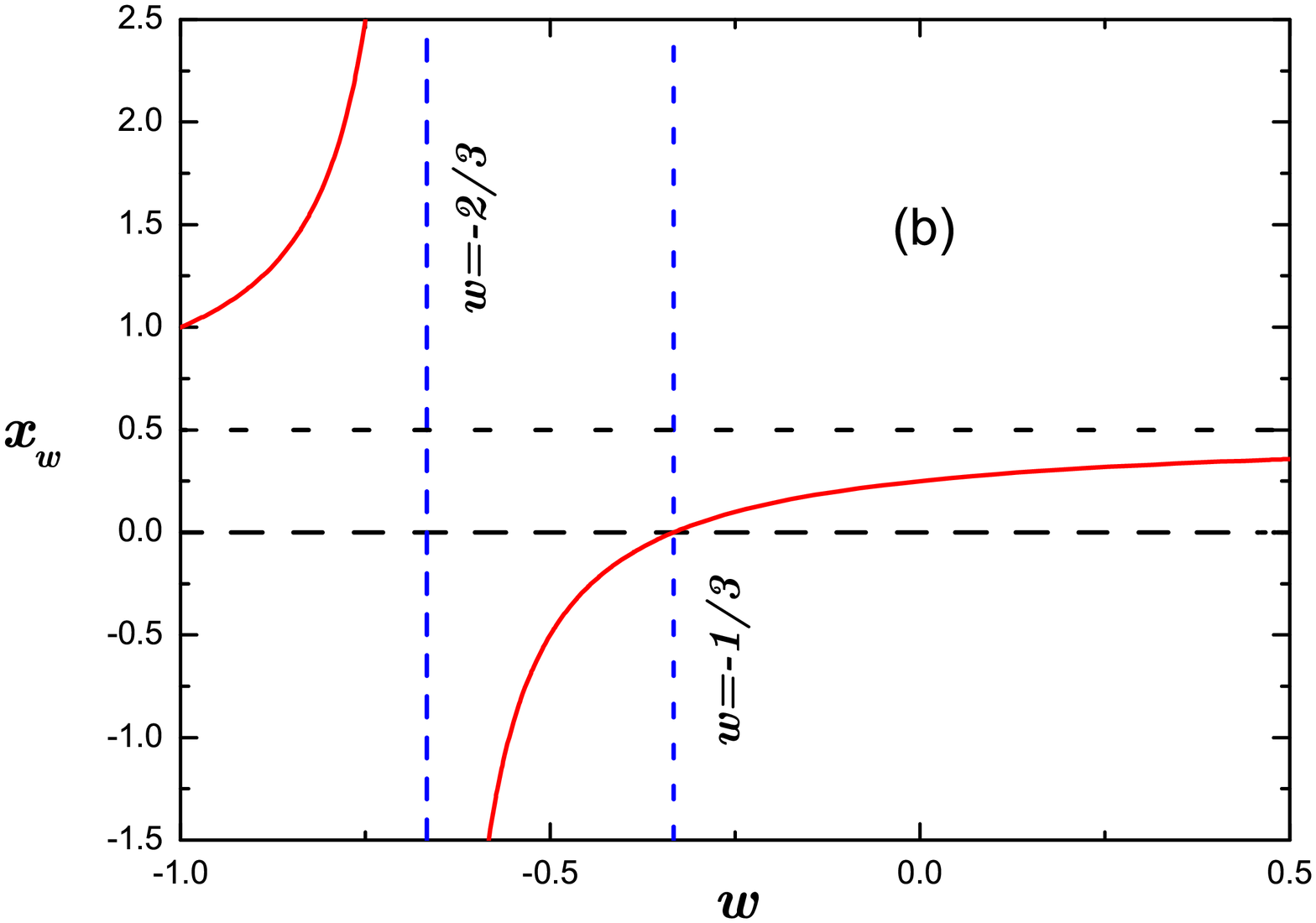}
\caption{(Color online)(a): The evolutionary picture of EoS of tachyon field with exponential potential and inverse quadratic one. The dot line is $\omega=-\frac13$. (b): the relationship between $\omega$ and $x_\omega=\frac{1+3\omega}{4+6\omega}$, we assume $\omega$ is a constant during the inflation stage.}\label{fig1}
\end{figure}

As a basis for the next section, we discuss the general properties of slow-roll inflation theory in LQC in this section. This job has been shown in \cite{Xiong,Xin}, but for completeness's sake, we briefly review the discussion of \cite{Xiong,Xin} in the following. To make the discussion of slow-rolling inflation easy, we usually introduce a slow-rolling parameter $\epsilon$ which is related to the evolution of the Hubble parameter
\begin{eqnarray}
\varepsilon=-\frac{\dot{H}}{H^2}=\frac32(1+\omega)\frac{1-2x}{1-x}
\end{eqnarray}
with $x=\frac\rho{\rho_c}$ and $\omega$ is the equation of state (EoS) of tachyon field
\begin{eqnarray}
\omega=\frac{p}{\rho}=\dot{\phi}^2-1. \label{omega}
\end{eqnarray}
The expression of EoS for tachyon field is only depends on its kinetic energy rather than its potential energy. But remember that the form of potential will effect the changing rate of kinetic energy, just as Eq.(\ref{EoM}) showed. According to Eq.(\ref{omega}), it easy to see $\omega\in [-1,0]$. We show the evolution picture of $\omega$ in Fig. \ref{fig1}(a). The tracks of EoS for two different potentials have similar trajectory at the very early time, and they become very different at the late time. But it is should be noticed that, the evolution of EoS depends on the initial values of the kinetic energy and potential energy of the tachyon field, and we will discuss these values in the next section.

The slow-rolling inflation happens while $\varepsilon<1$. The condition $\varepsilon<1$ is totally equivalent to the condition $\frac{\omega(1-2x)-x}{1-x}<-\frac13$. To be specific, it is $x>x_\omega=(1+3\omega)/(4+6\omega)$ for $\omega>-\frac23$, and $x<x_\omega$ for $\omega<-\frac23$ \cite{Wu}. But noticed that $x\in [0,1]$, therefore, the conditions for inflation are \cite{Wu}
\begin{eqnarray}\label{inc}
\left\{ \begin{aligned}
\frac12>x> x_\omega>0, &~~~ \text{for} & \omega>-\frac13,  \\
0  <x< \frac12, &~~~ \text{for} & -1<\omega<-\frac13.
\end{aligned} \right.
\end{eqnarray}
Assuming $\omega$ is a constant during the inflation stage, we give the relationship between $\omega$ and $x_\omega$ in Fig. \ref{fig1}(b).  For tachyon field, if the condition $\rho+3p<0$ is violated, i.e., $\omega<-\frac13$ is violated, the inflation will stop in Einstein cosmology. But just as we show in the Eq.(\ref{inc}), the inflation will not stop in LQC. This effect is caused by the quantum geometry. So, as a word, the quantum geometry changes the inflation in two aspects: one is the super inflation stage, which is totally cased by the quantum effect; the another one is inflation, the inflation still continues while $\omega\geq-\frac13$.

In this section, we give some basic equation for tachyon field in LQC, and discuss the general inflation theory for LQC, we will discuss the inflation theory for tachyon field with exponential potential and inverse quadratic potential in the next section.

\section{Tachyonic Inflation\label{III}}

In the last section, we showed the general inflation theory in LQC. In this section, we will discuss the super inflation for tachyon field with exponential potential and inverse quadratic one with different initial condition at first, and then, we will study the slow-roll inflation of tachyon field with exponential potential, and we will give more discussions about tachyonic inflation in LQC.

\subsection{Super inflation}

Since there is an additional $-\frac{\rho^2}{\rho_c}$ term in Friedmann equation (see Eq. (\ref{Fri})), the Hubble parameter should be zero when the energy density of tachyon field $\rho$ equals to the critical energy density $\rho_c$. At this moment, $\dot{H}>0$ for $(1-2\frac{\rho}{\rho_c})=-1$. This means the universe will enter an super inflation stage as soon as the bounce begins, and it will last until $\dot{H}=0$, i.e., $\rho=\frac12\rho_c$. There are lots of studies discuss the properties of super inflation \cite{Singh-PRD,Copeland-superinflation,Ranken,Ribassin,Amoros-prd,Xiao-Superinflation}. As we know, to get enough e-folding number, the scale factor should increase very fast in slow-roll inflation stage to ensure the e-folding number $N>60$.  According to the truth that the e-folding number comes from the changing of the scale factor in slow-roll inflation, some concluded that the e-folding number created during super inflation in LQC is not sufficient \cite{Singh-PRD}, but then it was found that the e-folding number during super inflation depends on the initial conditions  \cite{Mielczarek} or the parameter of the potential of scalar field \cite{Ranken}. Noticed that the Hubble parameter $H$ increases during super inflation stage, some argue that the super inflation can provide enough e-folding number for the expression of $N$ corrected as $\bar{N}\equiv \frac{a(t_f)H(t_f)}{a(t_i)H(t_i)}\to\infty$ \cite{Amoros-prd,Xiao-Superinflation}. In this subsection, we just consider the e-folding number from the change of the scale factor during the super inflation stage.

\begin{figure}[!ht]
\includegraphics[width=0.48\textwidth]{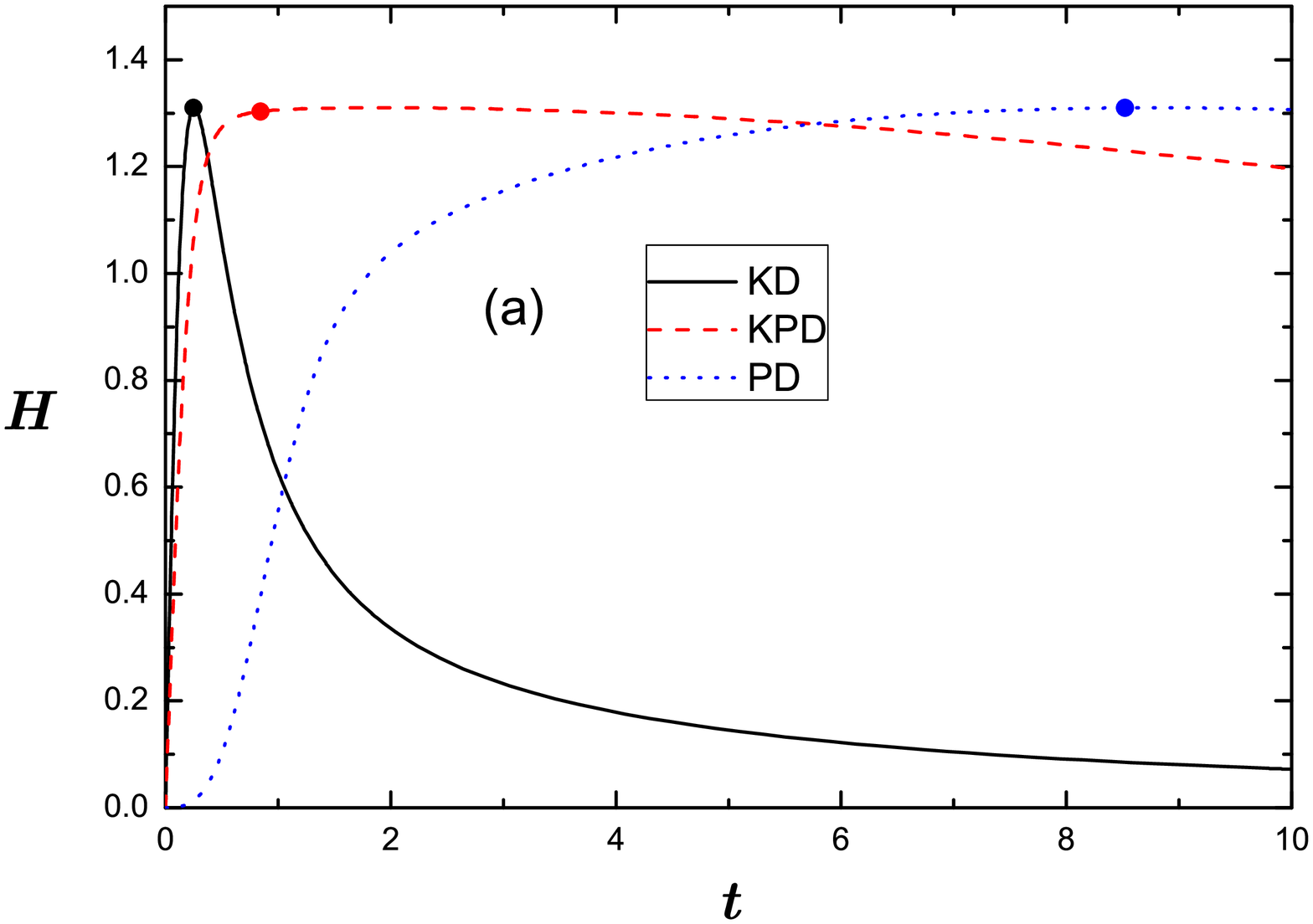}
\includegraphics[width=0.48\textwidth]{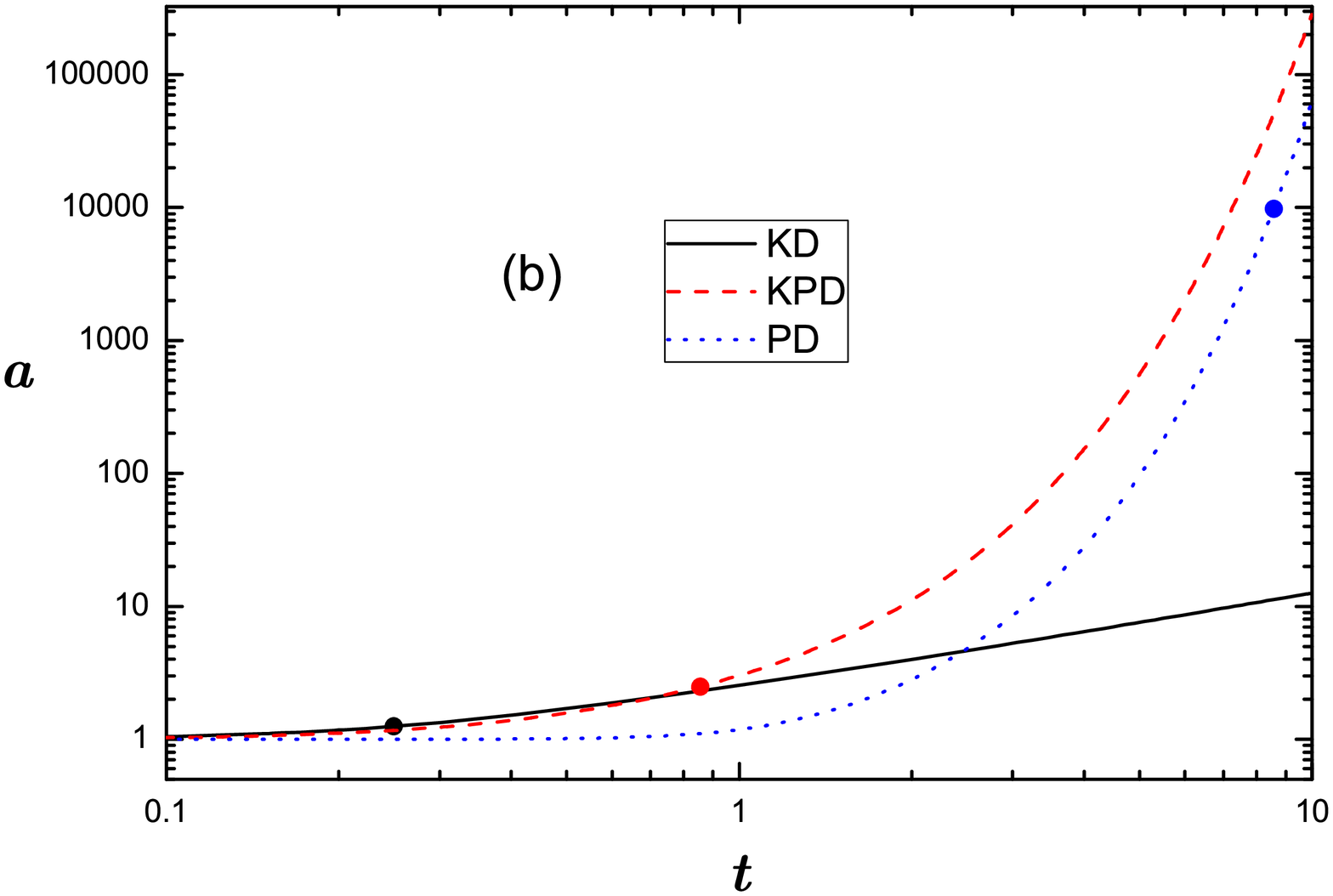}
\caption{(Color Online) The evolutionary pictures of the Hubble parameter $H$ and the scale factor $a$ with different initial conditions. The universe is soured by the tachyon field with the exponential potential.There are three kinds of initial conditions: KD: the kinetic energy dominates at the bounce point; KPD: the kinetic and potential energy sub-dominates at the bounce point, and PD: the potential energy dominates at the bounce point. The dot at the line represents the ending point of super inflation.}
\label{fig2}
\end{figure}

\begin{figure}[!ht]
\includegraphics[width=0.48\textwidth]{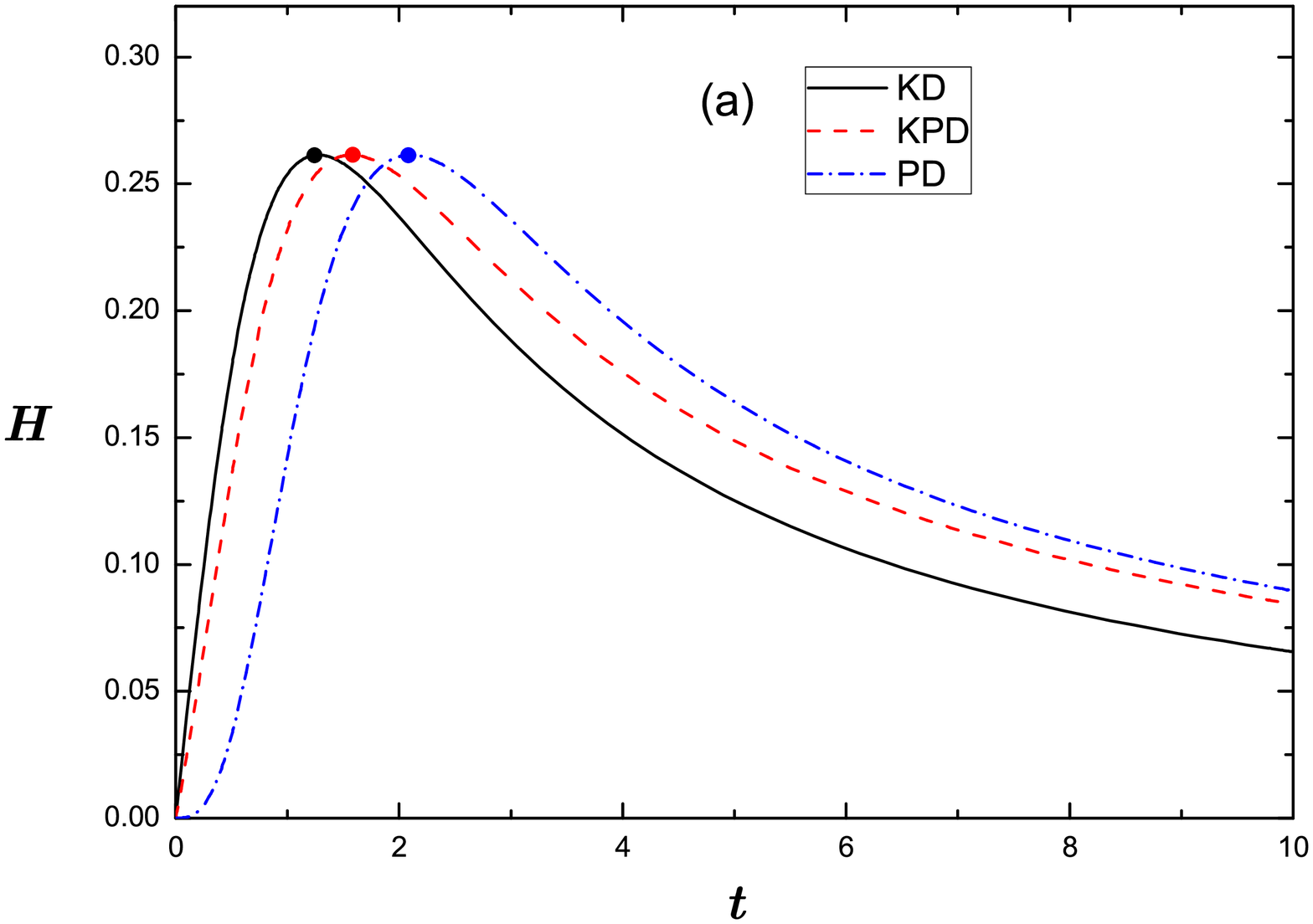}
\includegraphics[width=0.48\textwidth]{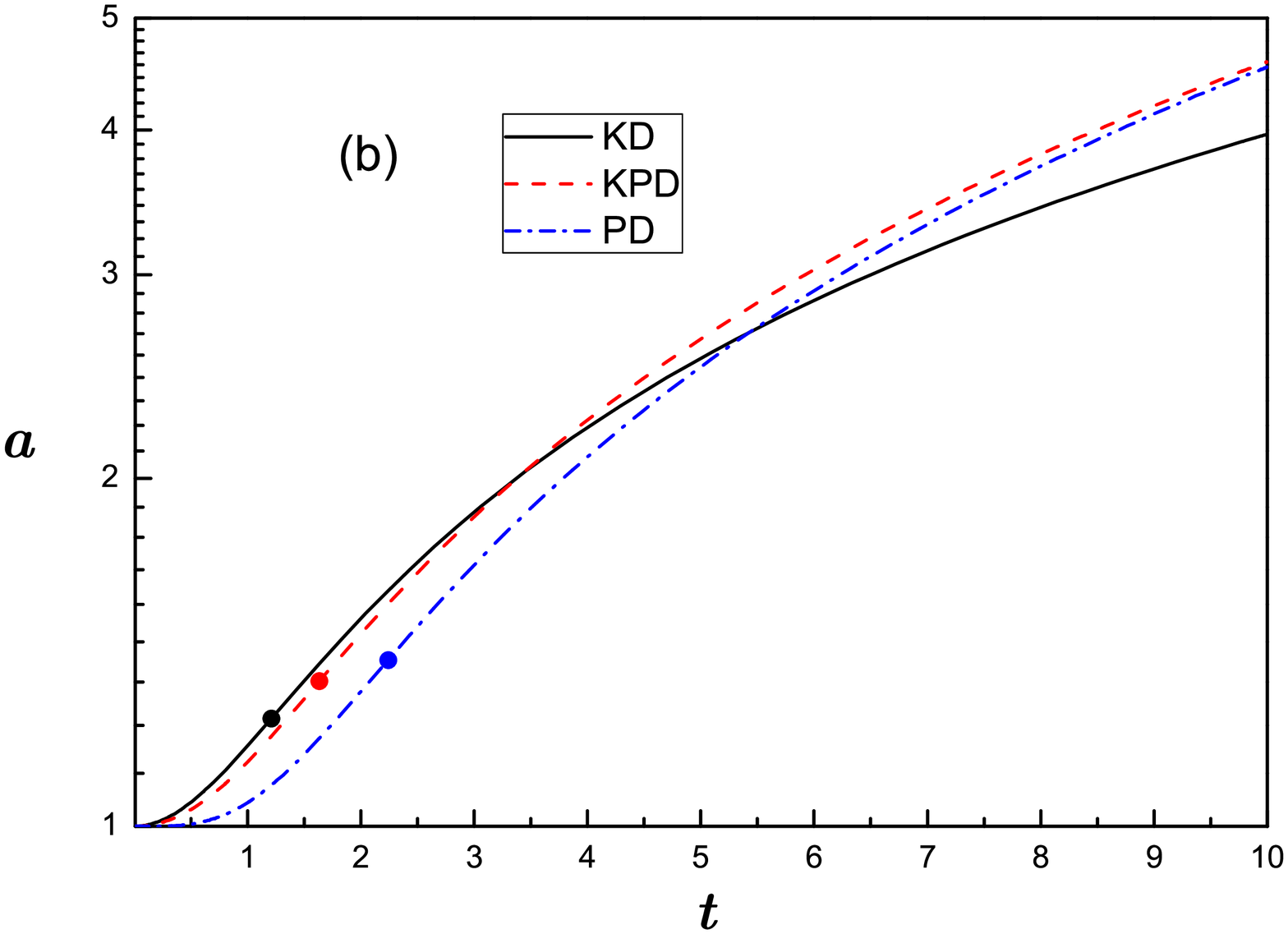}
\caption{(Color Online) The evolutionary pictures of the Hubble parameter $H$ and the scale factor $a$ with different initial conditions. The universe is soured by the tachyon field with the exponential potential.The three different conditions are as same as the ones in Fig. \ref{fig2}. The dot at the line represents the ending point of super inflation.}
\label{fig3}
\end{figure}

We discuss the super inflation of tachyon field with exponential potential and inverse quadratic potential. The purposes of this subsection are two: one is to find out whether the duration of super inflation for tachyon field depends on the potential, the other is to study whether the change of scalar factor depends on the initial values of kinetic and potential energy during super inflation.

We consider three kinds of initial conditions: I, the potential energy dominates (PD) at the bounce point; II, the kinetic energy and potential energy sub-dominates (KPD) at the bounce point ; and III, the kinetic energy dominates (KD) at the bounce point. The evolution pictures for Hubble parameter $H$ and scale factor $a$ for different initial conditions are shown in Figs. \ref{fig2} and \ref{fig3}. Fig. \ref{fig2} shows the evolution pictures for $H$ and $a$ under the condition that the universe is sourced by tachyon field with exponential potential, while Fig. \ref{fig3} shows that the evolution pictures for the tachyon field with inverse quadratic potential. We set the potential's parameter as a constant,  $V_0=0.82, \lambda=0.5, \beta=\sqrt{\frac53}$. These values just follows the choice of \cite{Xiong,Huang}.

The Hubble parameter $H$ increases immediately when the quantum bounce happens and it is lasts until $\dot{H}=0$, i.e., $\rho=\frac12\rho_c$, just as Fig. \ref{fig2} and \ref{fig3} shown. Also, it is easy to find out that the growth rates of $H$ are almost the same for the tachyon fields with the same potential but with different initial condition, and the growth rates are different for the different potential. This means that although the super inflation is cased by the quantum geometry effect, but the duration of it will be affected by the matter (i.e.,  same field with different potentials, or different fields). This result is tenable for interacting or non-interacting tachyon field \cite{Xiao}, as well as for the interacting or non-interacting scalar field \cite{Xiao-Superinflation,Ranken}. But the growth rates of scale factor $a$ are very different from the ones of Hubble parameter $H$. It is easy to find that the growth rates of $a$ are very different for the tachyon field with exponential potential under different initial conditions, but the difference is very small (at least in the same order of magnitude) for inverse quadratic potential.

To get enough e-folding number, if we consider the e-folding number just depends on the change of scalar field $N\equiv\ln\frac{a_f}{a_i}$, the scale factor needs to increase at least $e^{60}$ times. If one consider the universe sourced just by the tachyon field with the inverse quadratic potential, it is impossible to get enough e-folding number, just as Fig. \ref{fig3} showed, no matter which initial condition chosen. Although the scale factor $a$ increases very fast when one considers the initial condition that the potential energy dominates at the bounce point, the change rate of $a$ is still less than $e^{60}$. This result is different from \cite{Xiao-Superinflation}, in which we found that it is possible to get e-folding number during the super inflation of the interacting scalar field and radiation, i.e., the change rate of scale factor $a$ is bigger than $e^{60}$ if the potential energy dominates at the bounce point \cite{Xiao-Superinflation}.

In this subsection, we discussed the super inflation of tachyon field with exponential potential and inverse quadratic potential in LQC. We find that it is impossible to get enough e-folding number if we consider it comes from the change of the scale factor. To get enough e-folding number, if we still just consider the changes of the scale factor, the slow-rolling inflation is necessary. We will discuss the slow-rolling inflation in the next subsection.

\subsection{Slow-rolling inflation}

The initial condition for slow-rolling inflation is a hard nut to crack in Einstein cosmology, and the probability of inflation is suppressed by the factor $e^{-3N}$ \cite{Gibbons}, but the probability of inflation in LQC, with at least $N=65$ e-folding numbers, is very close to one for scalar field \cite{Ashtekar-probability-1}. The slow-rolling inflation for tachyon field in LQC has been discussed by many papers \cite{Xiong,Xiao,Wu}, we will discuss the tachyonic slow-rolling inflation in LQC based on these papers. As same as \cite{Xiong}, we discuss the tachyon with a exponential potential in this subsection.

\begin{figure}[!ht]
\includegraphics[width=0.48\textwidth]{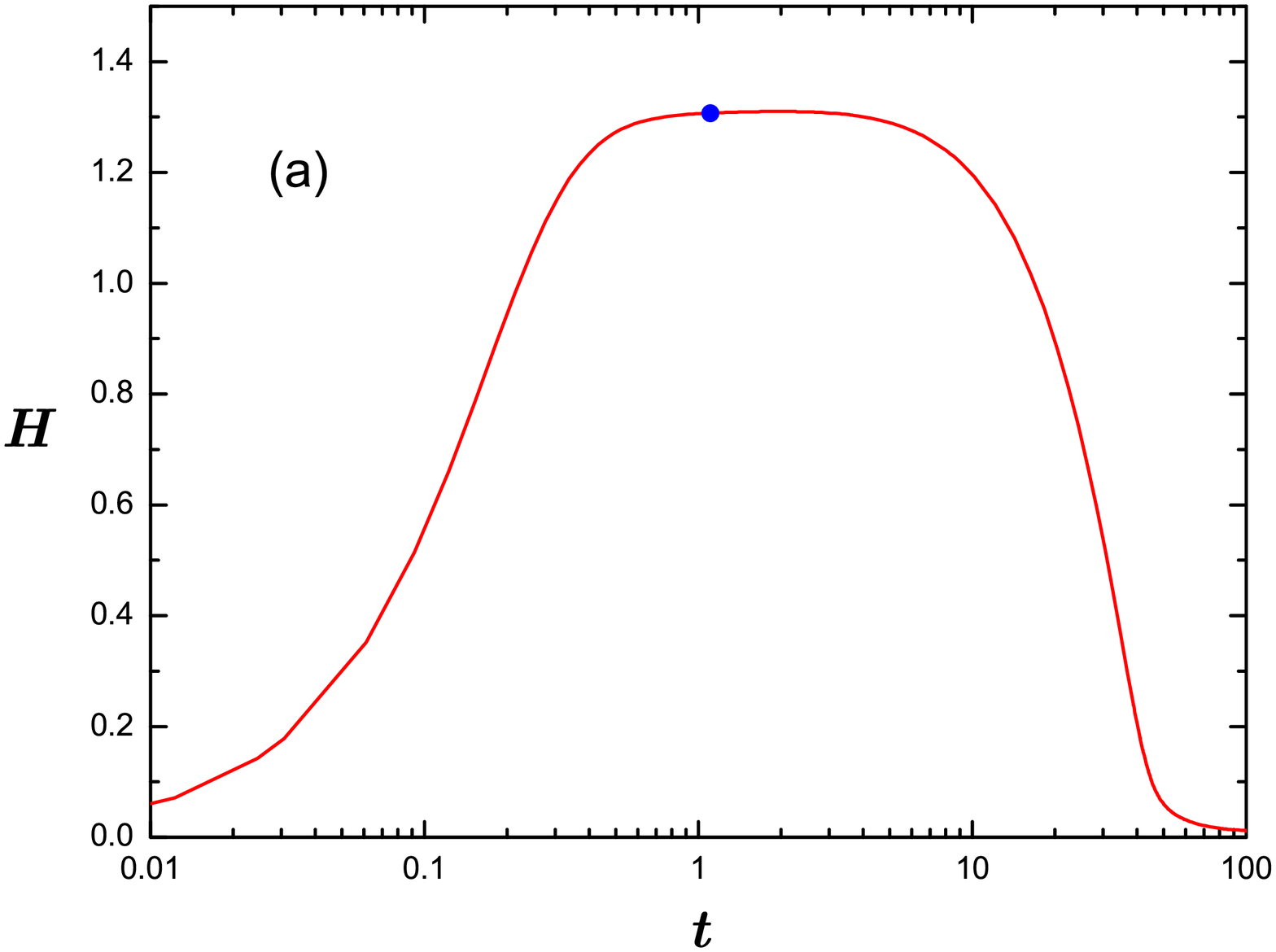}
\includegraphics[width=0.47\textwidth]{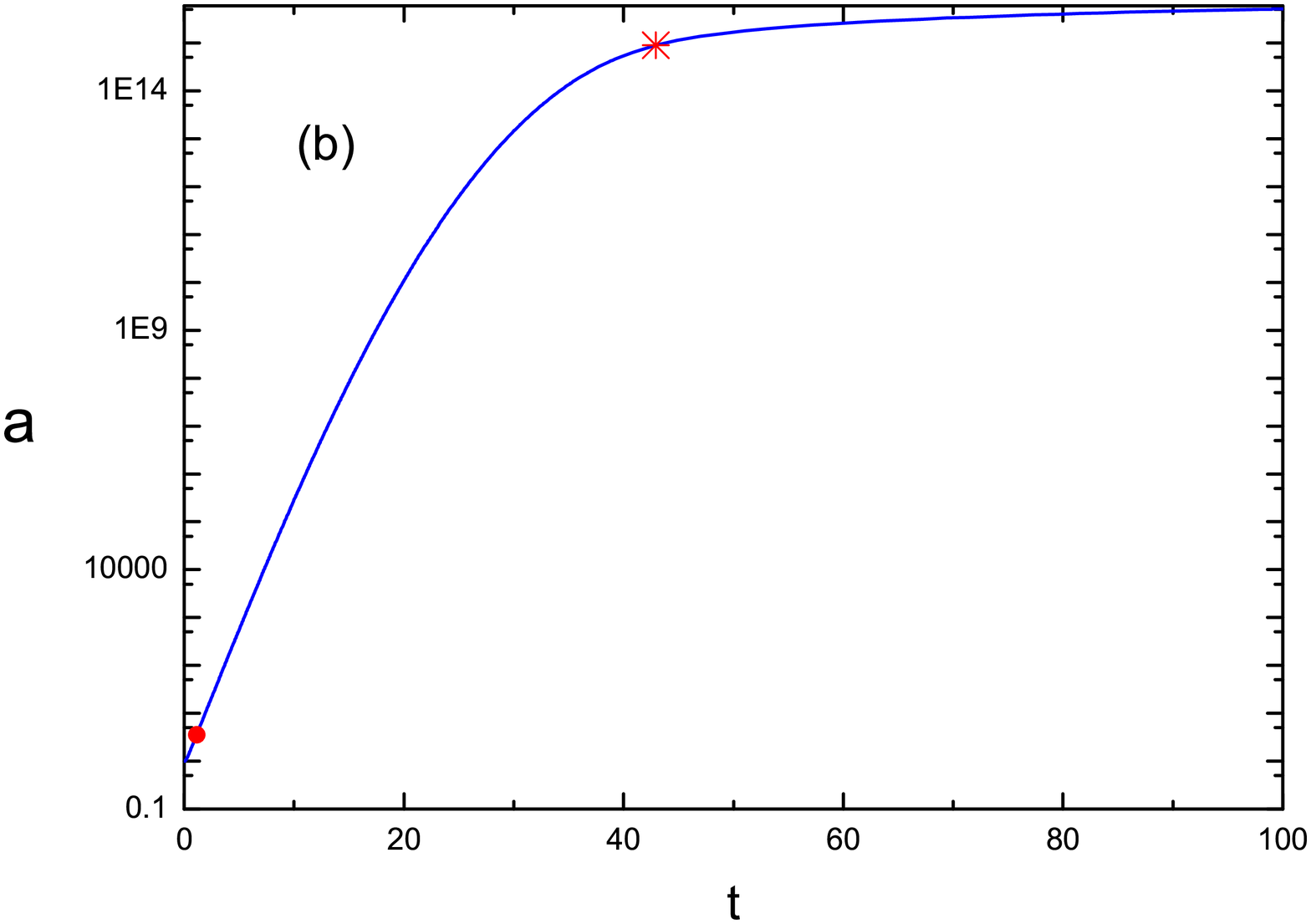}
\caption{(Color Online) The evolution pictures of the Hubble parameter $H$ and the scale factor $a$ for the condition that the kinetic and potential energy sub-dominates at the bounce point. The circle dot represents the ending point of super inflation, and the fork point denotes the end of the slow-rolling inflation.}
\label{fig4}
\end{figure}

\begin{figure}[!ht]
\includegraphics[width=0.48\textwidth]{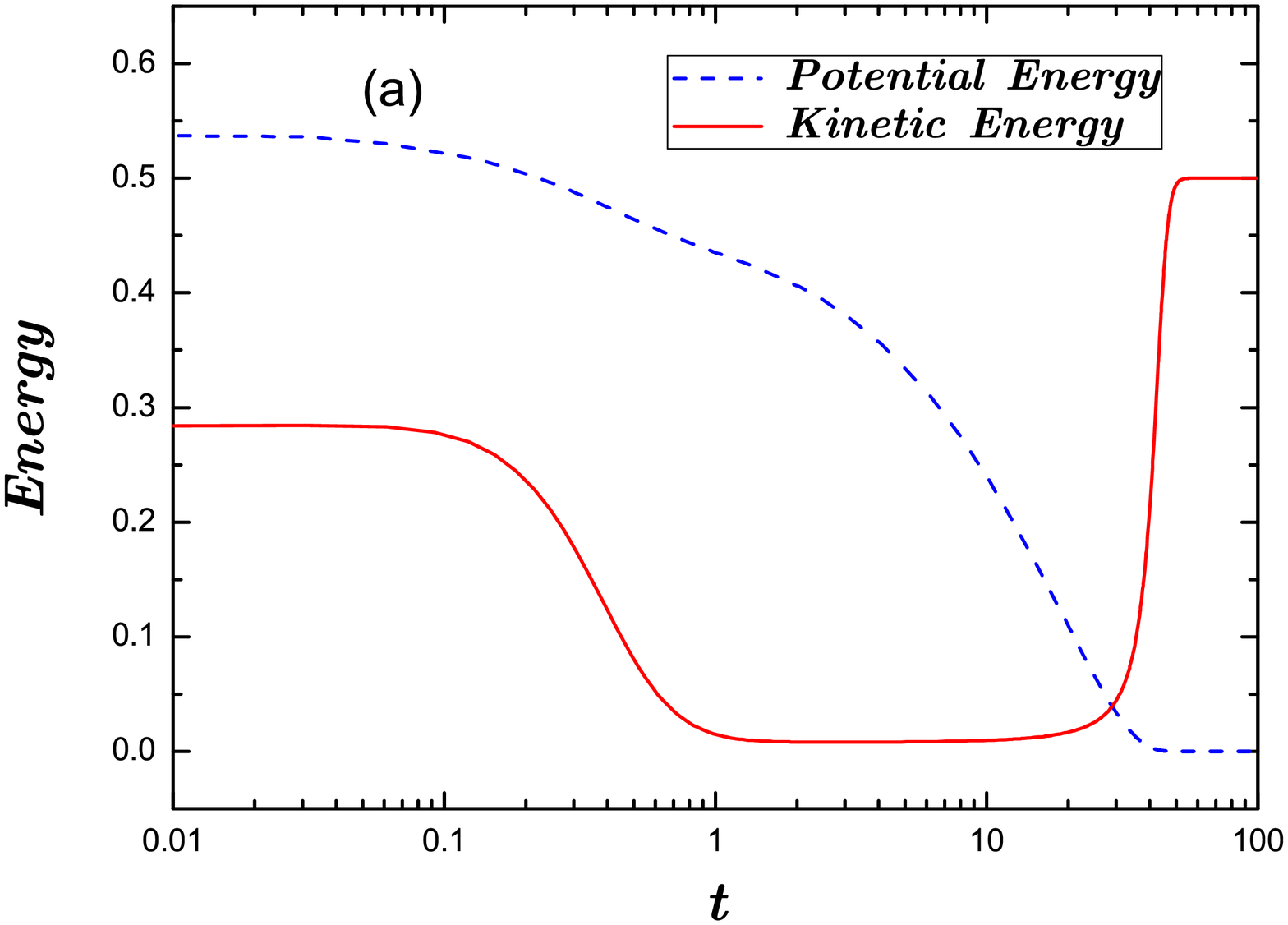}
\includegraphics[width=0.48\textwidth]{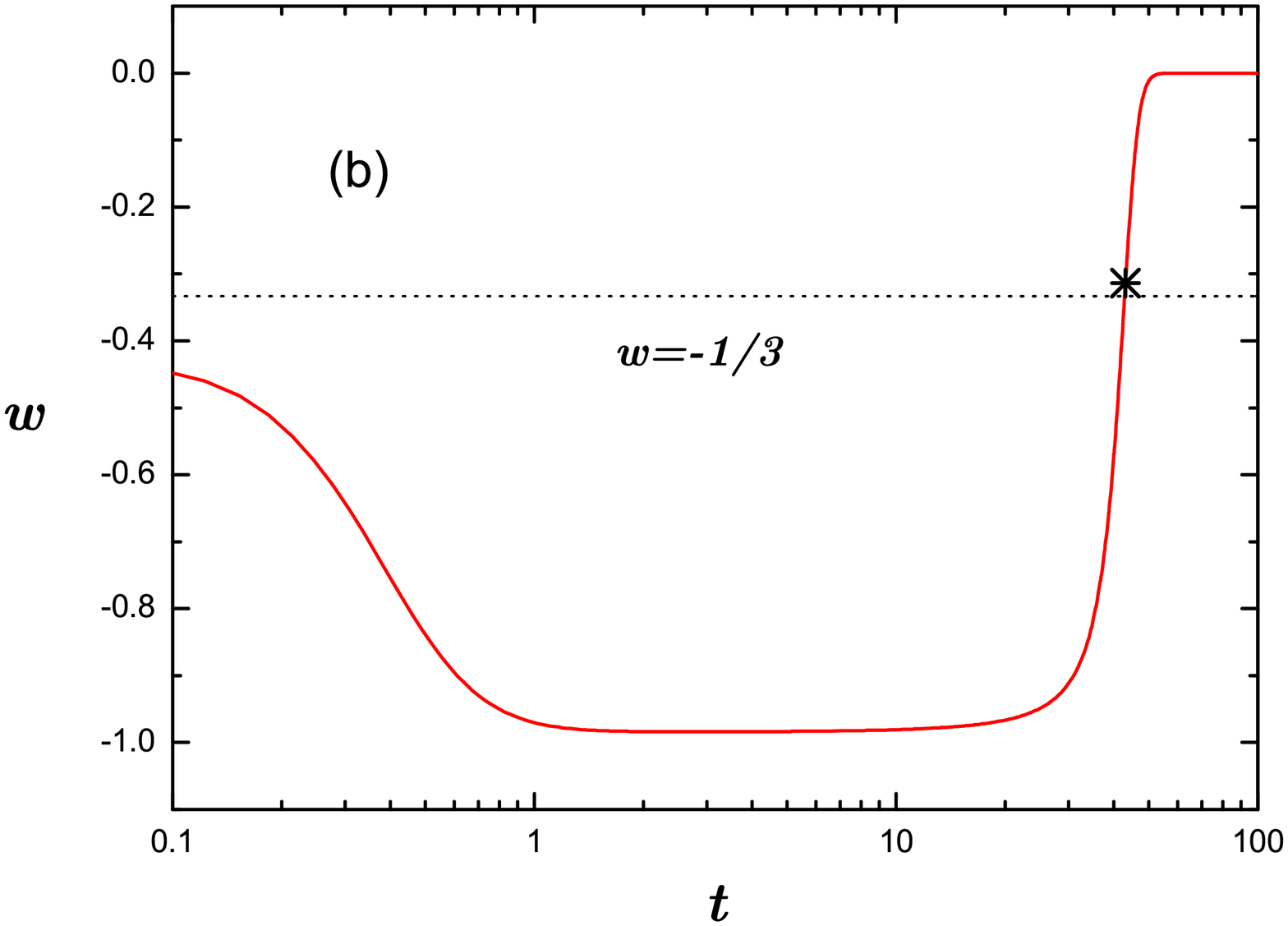}
\caption{(Color Online) (a): the evolution pictures of potential and kinetic energy. (b): the evolution picture for the EoS. The star dot denotes the end of the slow-rolling inflation. }
\label{fig5}
\end{figure}

The evolutionary pictures of the Hubble parameter $H$ and the scale factor $a$ are shown in Fig. \ref{fig4}. According to the numerical analysis, we find out that the Hubble parameter $H$ climbs up at the super inflation stage, holds the value for a while ( not a constant, but changes very slowly, just as the $H-t$ picture showed in Fig. \ref{fig2}), and then climbs down. But the scale factor changes rapidly. This appearance is just the requirement of the slow-rolling inflation in standard inflation theory. In the standard inflation theory, the second time derivative of $\phi$ is small enough, i.e., $|{\ddot{\phi}}|\ll |3H\dot{\phi}|, |V'|$, and the spacetime is approximately \emph{de sitter}, i.e., $a(t)\sim e^{Ht}$. The potential energy dominates the kinetic energy during inflation stage, this means that $\omega\sim -1$ at that time. We show the evolutionary pictures for potential energy and kinetic energy and the EoS in Fig. \ref{fig5}.
We find that the potential energy dominates the kinetic energy for a while, and the EoS hold $\omega=-1$ for a while and then increases to the value that the slow-rolling inflation ends (the star dot on the $\omega-t$ line in Fig. \ref{fig5}(b) and it persistently increases till $\omega=0$. Noted that the EoS $\omega\simeq-0.3319$ when the slow-rolling inflation ends in LQC. This value is bigger than $\omega=-\frac13$, and the slow-rolling inflation ending is $\omega=-\frac13$ in Einstein cosmology. This result is as same as the one we discussed in the last section, and it is another correction for inflation by the quantum geometry in LQC. Also, the track of $\omega$ in Fig. \ref{fig5} is different from the one in Fig. \ref{fig1}. This difference comes from the different initial conditions. All these signals show that the slow-rolling inflation happens as soon as the super inflation ends in LQC when we considers the potential and kinetic energy sub-dominates at the bounce point.

The slow-rolling inflation ends when the conditions is violated, i.e.,  $\varepsilon=1$. Unfortunately, there are still not enough e-folding numbers if the slow-rolling inflation ends when we chose $\alpha=0.5$ with the initial condition that the potential and kinetic energy sub-dominates at the bounce point. Just as the left picture shows in Fig. \ref{fig4}, the e-folding number in this case is $N\simeq 33$, less than $N=60$. To get enough e-folding number, a feasible method is choosing the parameters of potential very carefully, i.e., picking some smaller $\alpha$ \cite{Xiong}. Based on the method of \cite{Xiong}, and using the numerical analysis, we can find that the relationship between $\alpha$ and the e-folding numbers as showing in Table \ref{tab1}. The e-folding numbers $N$ is very different when one chooses the same $\alpha$ under different initial conditions. Obviously, the e-folding number is bigger when one chooses a smaller tachyon mass $\alpha$, no matter which initial condition is chosen. And the slow-rolling inflation is not happening for $\alpha=0.5$, and $0.6$ when one chooses the kinetic energy dominates at the bounce point, since the slow-rolling parameter $\varepsilon>1$ always.

\begin{table}[!ht]
\tabcolsep 3.2 mm
\begin{center}
\begin{tabular}{|c|c|c|c|c|c|c|}
\hline
$\alpha$&          0.1&0.2 & 0.3 &0.4 &0.5 & 0.6\\ \hline
$N_{\text{KPD}}$  & 763 &195& 106 & 57 & 33 & 20 \\ \hline
$N_{\text{PD}}$  & 768 &200& 87 & 47 & 31 & 21 \\ \hline
$N_{\text{KD}}$  & 680 &84& 18 & 7 &--& -- \\
\hline
\end{tabular}
\end{center}
\caption{ The relationship between the tachyon mass $\alpha$ and the e-folding number $N$. We set $V_0=0.82$. KPD,PD and KD respectively denotes the kinetic-potential dominates, potential energy dominates and kinetic energy dominates at the bounce point. ``--" denotes the slow-rolling inflation condition is violated. }\label{tab1}
\end{table}

In this subsection, we discuss the slow-rolling inflation for tachyon field with exponential potential in LQC. To get enough e-folding number, the tachyon mass $\alpha$ should be small, at least should be smaller than 0.4 for potential/kinetic-potential energy dominates at the bounce point and 0.3 for kinetic energy dominates at the bounce point. The slow-rolling inflation does not happen for all the times, i.e., at least, when $\alpha=0.5$ or $\alpha=0.6$, and the kinetic energy dominates at the bounce point. If one considers the tachyon field with inverse quadratic potential, it is easy to get similar conclusions, just as \cite{Xiong} argued.

\subsection{More discussions on tachyonic inflation}

In the last two subsections, we discuss the super inflation and slow-rolling inflation for tachyon field in LQC. We find that the e-folding numbers of super inflation is less than $60$, and whether the slow-rolling inflation provide enough e-folding number depends on the choice of tachyonic mass $\alpha$. To get enough e-folding number, one way is consider the e-folding numbers $\bar{N}\equiv \frac{a(t_f)H(t_f)}{a(t_i)H(t_i)}\to\infty$ during the super inflation stage for $H(t_i)\sim 0$ in the bounce point, just as shown in \cite{Amoros-prd,Xiao-Superinflation}; another way is choosing the tachyon mass very careful, and the slow-rolling inflation provides enough e-folding number, just as the choice of \cite{Xiong} and we showed it in the last subsection.

\begin{figure}[!ht]
\includegraphics[width=0.46\textwidth]{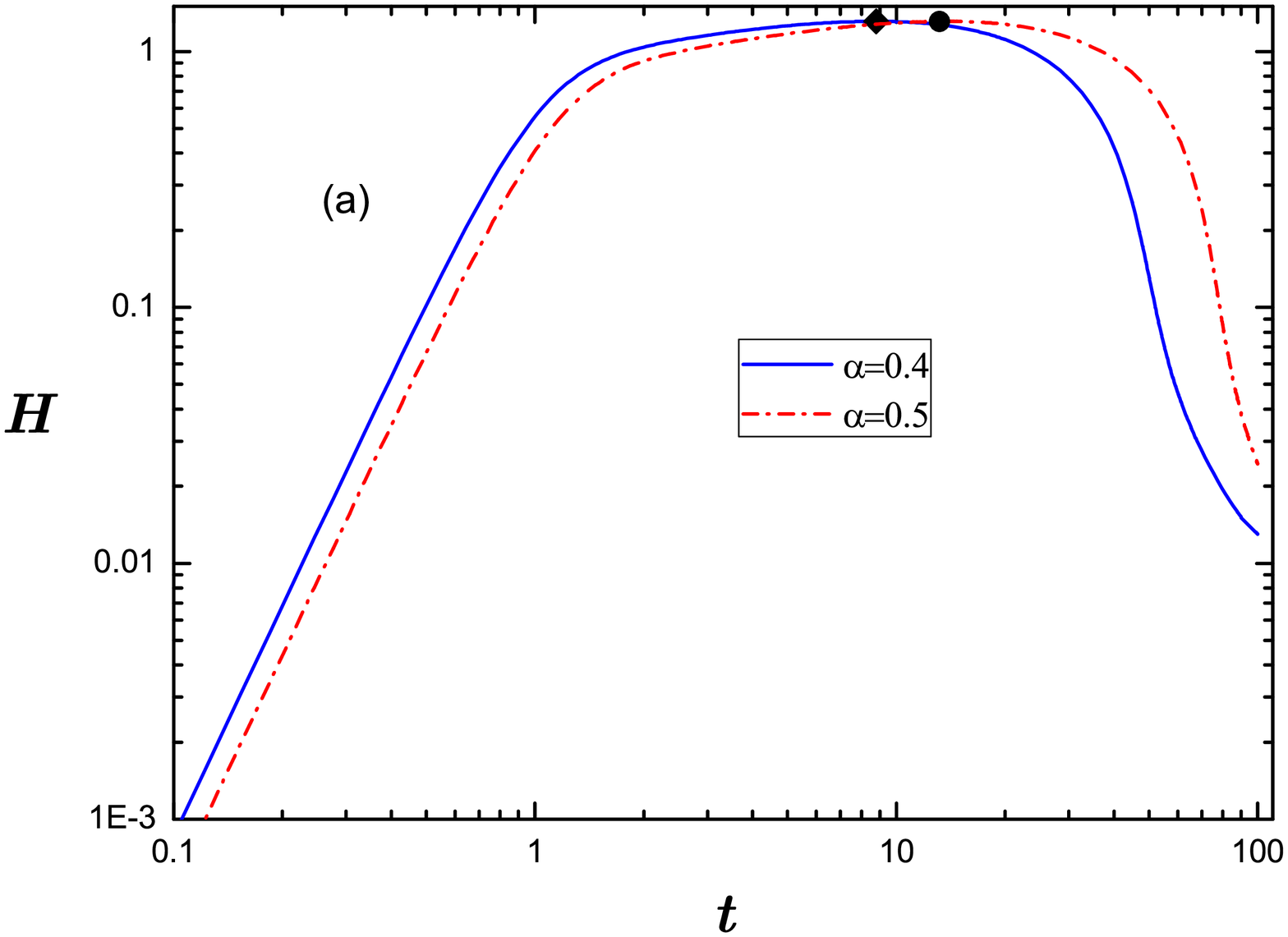}
\includegraphics[width=0.46\textwidth]{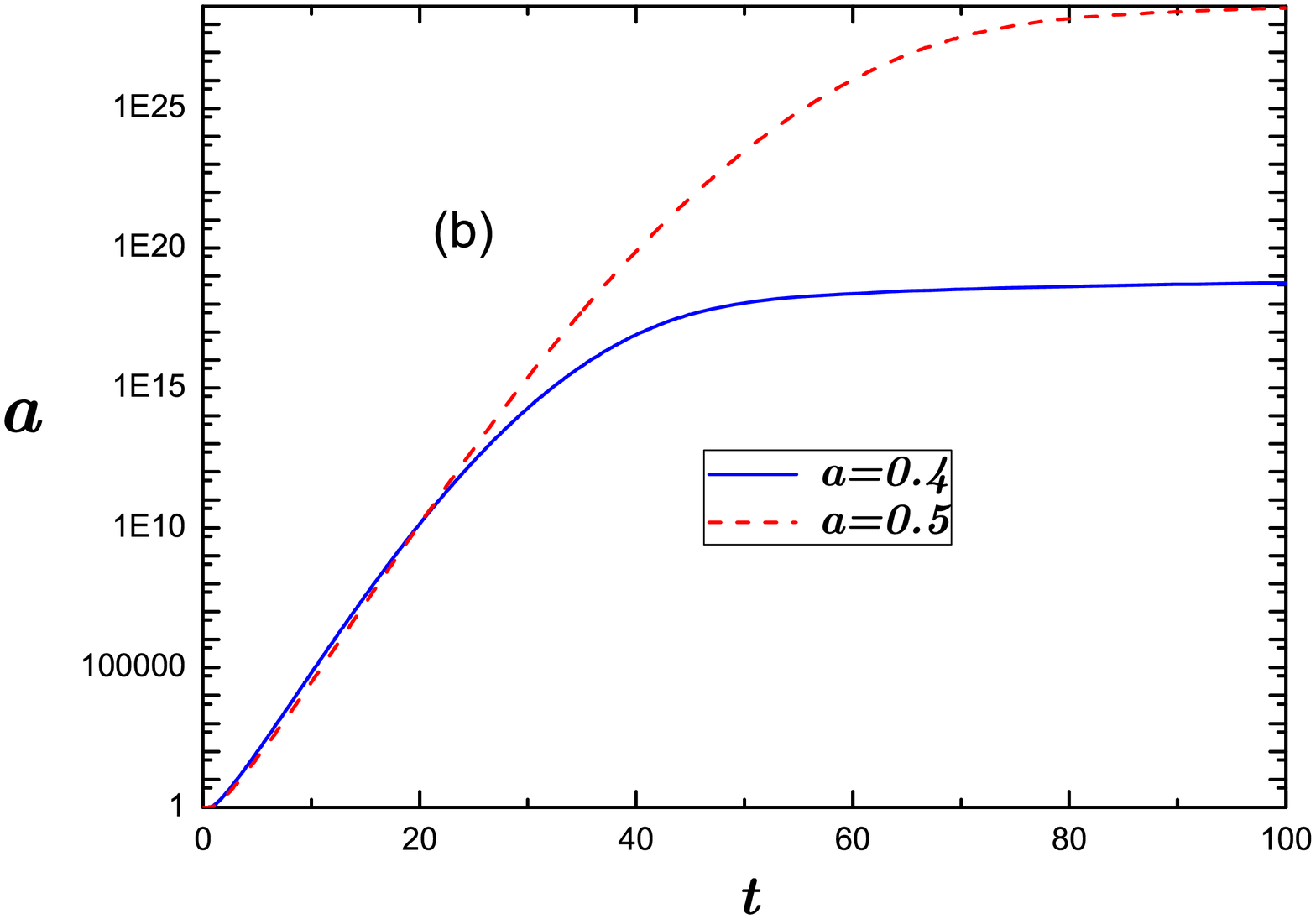}
\caption{(Color Online) Left: the evolution pictures of the Hubble parameter $H$. The circle dot in the  solid line and the trigonometric point in the dash dot line respectively denote  the ending point of the super inflation.  Right: the evolution picture for the scale factor. We consider the tachyon field with exponential potential, and the potential energy dominates at the bounce point.}
\label{fig6}
\end{figure}

Considering the e-folding number $N$ just depends on the changes of the scale factor, there is still one more question needs to be discussed. We show the evolution trajectory of the Hubble parameter $H$ in Fig. 6 (a). It is easy to find that the potential energy dominates the kinetic energy as soon as the super inflation ends and the EoS holds the value $\omega\simeq -1$ just we showed in the Fig. \ref{fig5}. Then, we can get the conclusion that the slow-rolling inflation happens as soon as the super inflation ends. This means that, considering the potential energy dominates at the bounce point, the inflation for tachyon field with exponential potential continuously happens. During the whole inflation stage, the Hubble parameter increases during the super inflation stage, the increasing rate is very small, and it almost a constant during the slow-rolling inflation stage. But the scale factor sustains the growth during the whole inflation stage, from the quantum bounce point to the ending time of slow-rolling inflation. This means that the e-folding number has two contributions, one from the increasing of the scale factor during the super inflation stage, and the another one from the changing of it in the slow-rolling inflation phase. Considering this truth, with the help of the numerical analyze, it is easy to get the total e-folding numbers during the whole inflation stage: $N_{\alpha=0.4}\simeq 65$ and $N_{\alpha=0.5}\simeq 41$. This means that it can provide enough e-folding number when the tachyon field mass $\alpha=0.4$.

The inflation in LQC has two phases, one is super inflation, and the another one is slow-rolling inflation. If the slow-rolling inflation happens as soon as the super inflation ends, we can treat this two inflation stages as a one inflation phase. During this inflation phase, the Hubble parameter $H$ increases at first and almost holds the values for a while until the inflation ends, and the scale factor $a$ becomes bigger and bigger during the whole inflation phase. In this inflation phase, the Hubble parameter $H$ increases very small, just as Fig. \ref{fig2} shown, but the change rates of scale factor $a$ can be very huge for some conditions. If one combines this two inflation stages as one inflation phase, it is easier to get enough e-folding number, just as we showed in this subsection, at least easier than just one considers the slow-rolling inflation.

A short conclusions for this section. We assume that the value of e-folding number just depends on the changes of the scale factor in this section. We find that the super inflation phase cannot provides enough e-folding number, and slow-rolling inflation with small tachyonic mass $\alpha$ maybe can solve it. But for the slow-rolling inflation happens immediately after the super inflation phase ends, the e-folding number is easier to provide for the scale factor increases during all the inflation phase.

\section{Conclusions and Discussions}

We discuss the tachyonic inflation in LQC in this paper. The inflation in LQC has two stage, one is the super inflation stage, which is totally caused by the quantum geometry effect, and the another one is the slow-rolling inflation stage. The slow-rolling inflation in LQC is different from the one in Einstein cosmology, for the existing of the quantum correction, just as the Eq. (\ref{inc}) shows.

Considering the tachyon field with exponential potential or inverse quadratic potential, we discuss  the super inflation for tachyon field in LQC. We consider three different initial conditions for two potentials, and get the evolutionary pictures for the Hubble parameter $H$ and the scale factor $a$. We find that, although the super inflation is a quantum geometry effect, the duration of super inflation is influenced by the matter. Considering the e-folding number is totally depending on the change of the scale factor, we find that the super inflation for tachyon field cannot provide enough e-folding numbers, neither for the tachyon field with exponential potential, nor for the one with the inverse quadratic potential, and the slow-rolling inflation is needed.

As an example, we discuss the slow-rolling inflation of tachyon field with exponential potential. After the super inflation, the universe enters a slow-rolling inflation immediately for the initial conditions that kinetic-potential energy dominates with $\alpha=0.5$ or potential energy dominates with $\alpha=0.4,0.5$ at the bounce point, just as Figs. \ref{fig4} and \ref{fig6} showed. To get enough e-folding numbers, the tachyonic mass $\alpha$ should be small. No matter which initial condition is chosen, the e-folding number is bigger for smaller $\alpha$.  And for some values of $\alpha$, the slow-rolling inflation won't even happen. The slow-rolling inflation happens while the super inflation ends, and the scale factor continuously increases during the whole inflation stage, then it is possible that the e-folding number comes from the changes of scale factor during the super inflation and slow-rolling inflation. Considering the potential energy dominates at the bounce point, we find that the total e-folding number for $\alpha=0.4$ is enough, but it is not enough for $\alpha=0.5$. Considering the super inflation stage and slow-rolling stage as a whole inflation phase, we find that it is easier to get enough e-folding number.

In this paper, we consider the e-folding number that comes from the changes of the scale factor, no matter this changes comes from the super inflation stage, or the slow-rolling inflation stage. Although the enough e-folding number can be provided in the super inflation stage \cite{Amoros-prd,Xiao-Superinflation}, but the slow-rolling inflation still needs to be studied in LQC. On the one hand, the slow-rolling inflation will be more likely to occur in LQC, just as we showed in the last section, and showed by many other papers  \cite{Singh-in,Ashtekar-probability-1,Ashtekar-probability,Corichi_measure,Linsefors,Singh-in-2}, on the other hand, the observational effect of super inflation is still absent, at least still needs to be studied \cite{yue}, but the observation issue of slow-rolling inflation in LQC has been discussed by many papers (see \cite{Barrau} for a review). Always, the observational data give some constraints on the potential and the parameters of the potentials. There are two parameters $V_0,\alpha$ for exponential potential and one parameter $\beta$ for inverse quadratic potential, we just consider there are some constants in this paper, but to find which values are suited, more study is needed, and the observational data should also be considered. We get a conclusion that the e-folding number is easier to get when the e-folding number depends on the changes of scale factor during the super inflation and the slow-rolling inflation phase if the slow-rolling inflation happens as soon as the super inflation ends. If this conclusion is correct, will the observational effects be evident? To understand the inflation theory more deeply in LQC, and know which parameter are suitable, we need consider the observational effect, just as \cite{Barrau,Haro-BICEP2} did. But the perturbation theory of tachyon matter in LQC is still empty, thus we just get some theoretical results in this paper.

\acknowledgments Zhu was supported by the National Natural Science Foundation of China (Grant Nos. 11175019 and 11235003), Xiao was supported by the National Natural Science Foundation of China (Grant No. 11175019 and 11401199), and He was supported by the National Natural Science Foundation of China (Grant No. 11401199).

\end{document}